\pdfoutput=1 
\documentclass[aps,prl,showpacs,twocolumn,groupedaddress]{revtex4}
\usepackage[pdftex]{graphicx}
\usepackage{amssymb}
\usepackage{amsmath}
\usepackage{psfrag}
\usepackage{epsfig}
\usepackage{hyperref}
\providecommand{\openone}{\leavevmode\hbox{\small1\kern-3.8pt\normalsize1}}

\newcommand{\tr}{\mathop{\mathrm{tr}}\nolimits}

\newcommand{\beq}{\begin{equation}}
\newcommand{\eeq}{\end{equation}}
\newcommand{\be}[1]{\begin{equation}\label{#1}}
\newcommand{\ee}{\end{equation}}

\newcommand{\bea}{\begin{eqnarray}}

\newcommand{\eea}{\end{eqnarray}}

\newcommand{\refeq}  [1] {(\ref{#1})}

\newcommand{\refFig} [1] {Fig.~\ref{#1}}

\begin{document}

\title{Observation of Periodic Orbits on Curved Two--dimensional Geometries}

\author{M.~Avlund$^{1}$, C.~Ellegaard$^{1}$, M.~Oxborrow$^{2}$, T.~Guhr$^{3}$,
        and N.~S\o ndergaard$^{4}$}

\affiliation{
$^1$Niels Bohr Institutet, Blegdamsvej 17, 2100 K\o benhavn \O, Denmark\\
$^2$National Physical Laboratory, Hampton Road, Teddington, TW11 0LW, UK\\
$^3$Fakult\"{a}t f\"{u}r Physik, Universit\"at Duisburg--Essen, Lotharstra\ss e 1
     47057 Duisburg, Germany\\
$^4$Matematisk Fysik, LTH, Lunds Universitet, Box 118, 22100 Lund, Sweden}

\date{\today}
 
\begin{abstract}
  We measure elastomechanical spectra for a family of thin
shells. We show that these spectra can be described by a 
``semiclassical'' trace formula comprising periodic orbits on
geodesics, with the periods of these orbits consistent with
those extracted from experiment. The influence of
periodic orbits on spectra in the case of two-dimensional
curved geometries is thereby demonstrated, where the parameter
corresponding to Planck's constant in quantum systems 
involves the wave number and the curvature radius. We
use these findings to explain the marked clustering of
levels when the shell is hemispherical.
\end{abstract}
\maketitle

When studying spectra of quantum systems, semiclassics, in particular
periodic orbit theory, provides a powerful connection to the dynamics
of the analogous classical system~\cite{gut90,bra97,sto99,haa01}.
Considerable progress was recently made in understanding generic
statistical properties of quantum chaotic
systems~\cite{heu07}. Furthermore, spectra of specific systems can be
constructed from periodic orbits, prominent examples are the Hydrogen
atom in a strong magnetic field~\cite{fri89}, the Helium
atom~\cite{win92} as well as regular and chaotic two--dimensional
billiards~\cite{bra97,sto99}. Microwave experiments in flat
cavities~\cite{mc2b} are most useful because, in two dimensions, the
Helmholtz equation for the electrical field formally coincides with
the Schr\"odinger equation.

Do the concepts of quantum chaos carry over to other wave phenomena? 
--- Elasticity, i.e.~mechanical vibrations, is a particularly
interesting testing ground. The governing equations and the boundary
conditions are different from the ones in quantum mechanics. Moreover,
different modes (pressure and shear) are present and propagate with
different velocities, in case of anisotropy even depending on the
directions. Hence, a transfer of quantum chaos ideas to elastic
systems is a worthwhile endeavor in its own right. The bulk waves in
three--dimensional aluminum blocks have the same statistical features
as known for quantum chaotic systems~\cite{ac3b}. Remarkably, even
much more subtle features have been measured and understood in a
framework transferred from quantum chaos: parametric
statistics~\cite{ber99} , transport and localization properties in
three--dimensional blocks~\cite{wea00}, the statistics of elastic
displacements, i.e.~``wave functions'', in two--dimensional
plates~\cite{sch03} and Wannier--Stark ladders in
quasi--one--dimensional elastic systems~\cite{gui06}.

In this contribution, we measure spectra of thin elastic shells and
explain certain characteristics by periodic orbits. This is, to the
best of our knowledge, the first experimental identification of
periodic orbit features on curved shells.  We have two goals. First,
we show that periodic orbit theory can be applied to the shells. This
is non--trivial because of the curvature and because the modes are a
combination of flexural and in-plane fields. They are in general
described by a system of partial differential equations of high order
\cite{kra67}, very different from the Schr\"odinger equation which is
the starting point in quantum chaos. Second, we will use this insight
to explain a striking clustering effect in the spectra. These findings
are not only of conceptual but also of practical importance. Metallic
shells are ubiquitous in technology and every day life, ranging from
auto bodies to micro--electro mechanical systems. Since the
calculation of spectral features from the wave equation is often
tremendously complicated, an understanding in terms of simpler
geometric quantities, the periodic orbits, might be of considerable
interest.

\begin{figure}[htp]
\begin{center}
\includegraphics[width=80mm]{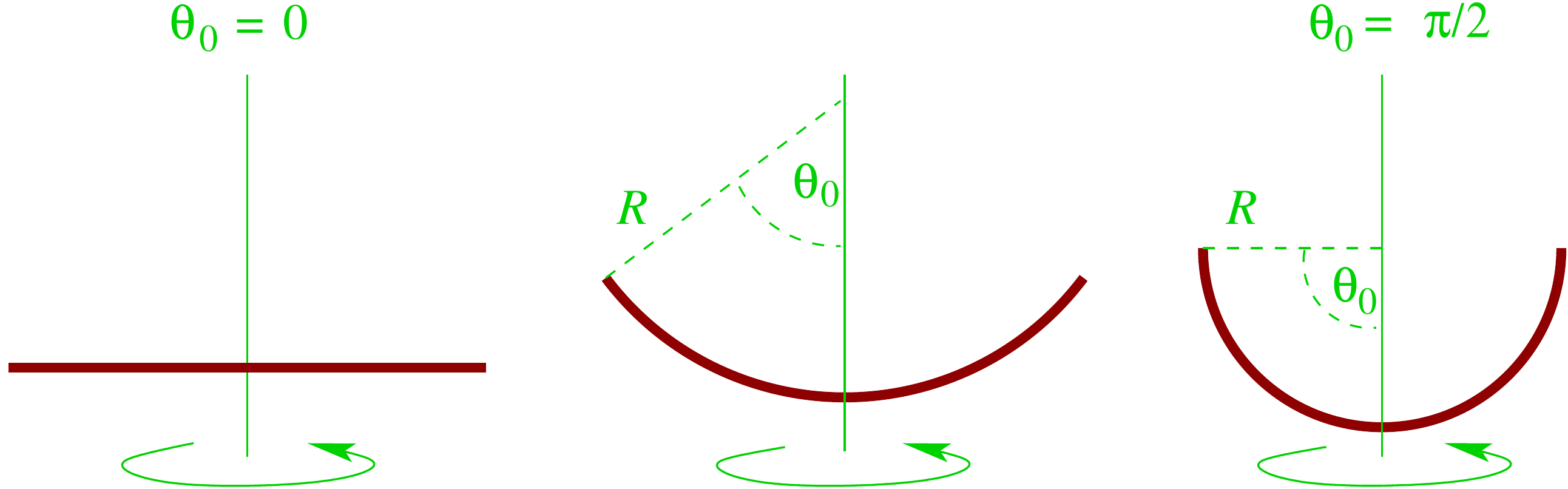}
\end{center}
\caption{(Color online) Shells of revolution with opening 
  angle $\theta_0$ and curvature radius $R$.}
  \label{fig1}
\end{figure}

\begin{figure}[htp]
\begin{center}
\includegraphics[width=80mm]{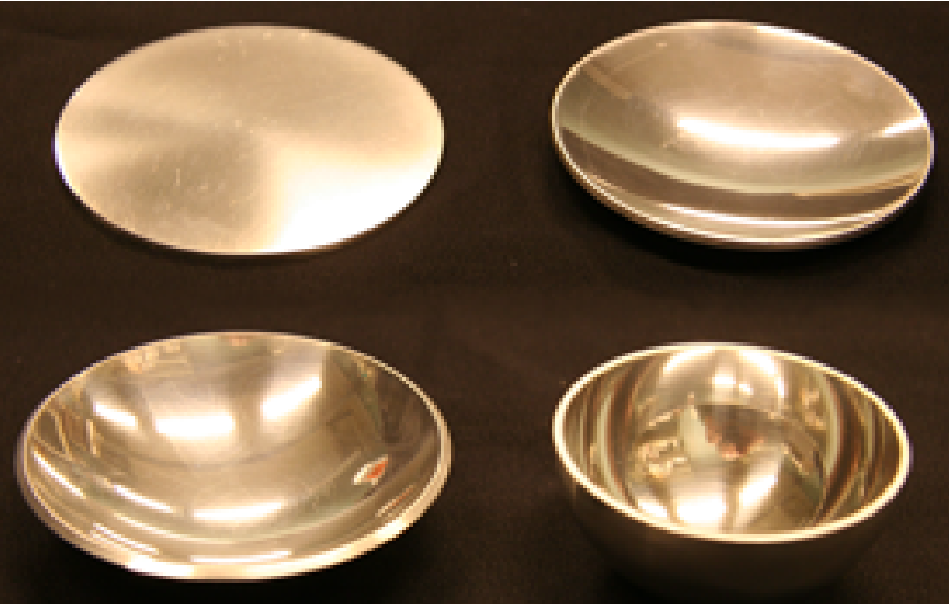}
\end{center}
\caption{(Color online) Aluminum shells used in the experiment.  From
  left to right, the opening angle is $\theta_0=0^\circ, 25^\circ,
  52^\circ$ and $90^\circ$.}
  \label{fig2}
\end{figure}

The shells we employ are objects of revolution as shown in
Fig.~\ref{fig1}. They form a family of constant mid-surface area parameterized by the opening
angle $\theta_0$ measured from the axis of revolution, whereas  $R$  is the   
$\theta_0$ dependent curvature. For $\theta_0=\pi/2$, we have a
hemisphere, the angle $\theta_0=0$ formally corresponds to the planar
disk. 
Figure~\ref{fig2}
displays the aluminum shells used in the experiment, the opening
angles are $\theta_0=0^\circ, 25^\circ, 52^\circ$ and $90^\circ$. Each
shell is 2~mm thick although for the hemisphere three thicknesses were studied.  The disk has a diameter of 80~mm, the other
shells were made in such a way that all mid--surface areas are equal.
The shells were carved from a solid aluminum block to avoid any
internal strain which inevitably occurs when producing curved objects
by bending a plate. It was carefully made sure that the cut defining
the boundary was along the radius, as seen in Fig.~\ref{fig1}. This is
important to prevent coupling between flexural and membrane--like
modes.

\begin{figure}[htp]
\includegraphics[angle=0,width=85mm]{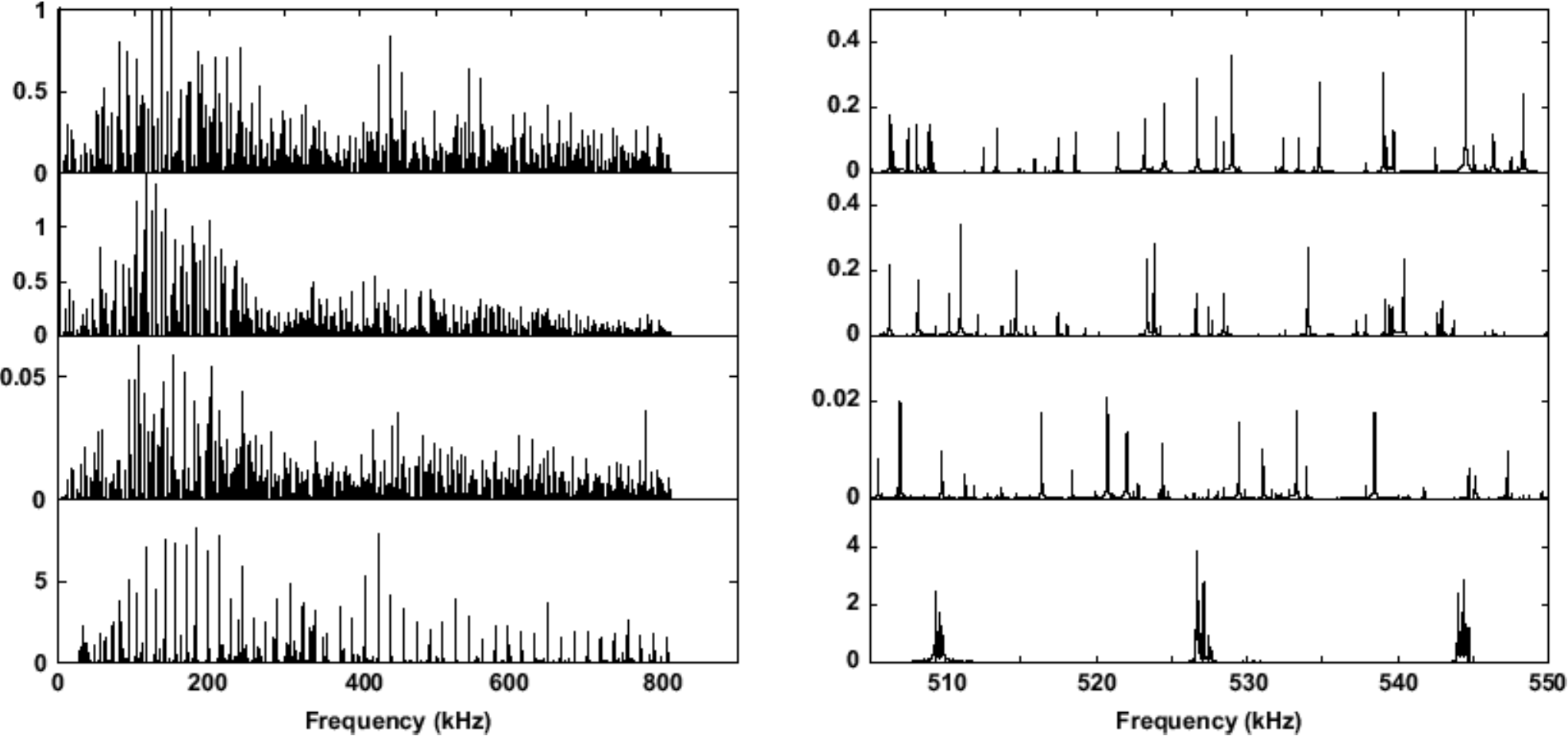}
\caption{(Color online) Measured spectra for the shells with
  opening angles $\theta_0=0^\circ, 25^\circ,
  52^\circ$ and $90^\circ$   from top to bottom.}
  \label{fig3}
\end{figure}

We used the experimental setup developed in our previous
studies~\cite{ber99,sch03}. It has a very high resolution, the quality
factor $Q=f/\Delta f$ is typically about $10^5$ where $f$ and $\Delta
f$ are position and width in frequency of a given resonance,
respectively. We accumulated data in the frequency range 0~kHz$\ \le f
\le \ $800~kHz. In the measurement, the shells are only supported by
three $1$ mm ruby spheres which minimize the elastic coupling to the
rest of the setup. The acoustic coupling to the air is significantly
reduced by putting the setup into a vacuum chamber whose inside is
held at $10^{-3}$ Torr. Flexural (bending) modes and membrane--like
(stretching and shearing) modes are excited.  The resulting spectra
are displayed in \refFig{fig3}. A visual inspection immediately
reveals that the
spectra for the disk and the shells with opening angle
$\theta_0<\pi/2$ look rather similar. For the hemisphere with
$\theta_0=\pi/2$, however, the levels are structured in clusters which
are almost equally spaced.  The clustering was already found in
Ref.~\cite{nio88} by numerically solving the equations for thin
shells~\cite{kra67}.  In the following, we will use periodic orbits to
give a clear and intuitive explanation of this striking effect.

Our reasoning will be in the spirit of semiclassical analysis for
quantum systems. We demonstrate that the \hbox{in-plane} excitations can be
described by a sum over periodic orbits, i.e.~by a trace formula akin
to the ones in quantum chaos~\cite{gut90,bra97,sto99,haa01}. This goes
well beyond previous work on the celebrated whispering gallery
modes~\cite{bra97}, on the ray--description of seismic
waves~\cite{dah98} as well as on the identification of periodic orbits
in elastic spectra of three dimensional systems~\cite{del94,ber99}.
\begin{figure}[htp]
\begin{center}
\includegraphics[width=80mm]{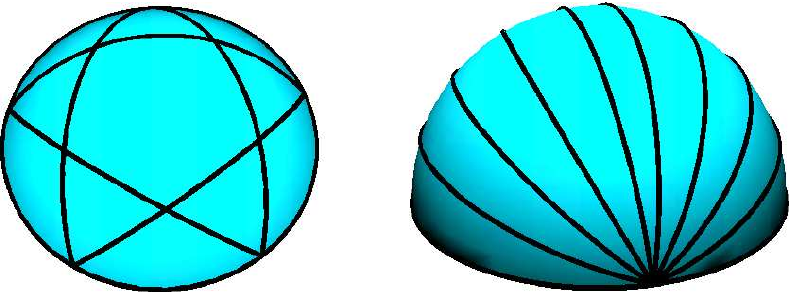}
\end{center}
\caption{(Color online)  Geodesic pentagram and diameters on
  the spherical cap.}
  \label{fig4}
\end{figure}
Trace formulae are different for regular and chaotic systems. If the
wave equation describing shell vibrations were scalar, all our shells
would be integrable. The complexity of the wave equation for thin
shells~\cite{kra67} modifies that picture. The wave equation can be
separated into an angular part depending on the angle of revolution
and a ``radial'' part depending on the angle $\theta$ measured from
the axis of revolution with $0 \le \theta \le \theta_0$. In the
``semiclassical'' approximation~\cite{son08} the waves are
approximated by the motion of a fictitious particle. The main insight
is that this motion takes place on the geodesics of the shells as in
\refFig{fig4}. The r\^ole of Planck's constant compared to a typical
action is here played by $1/kR$, where $k$ is the wave number.  To
leading order, the flexural and the membrane--like motion decouple,
i.e.~yield two separate equations of motion for the fictitious
particle. Hence, the flexural motion is integrable. There are two
degrees of freedom on the shells and two constants of motion, the
energy and the angular momentum with respect to the axis of
revolution. The membrane--like motion, however, has two polarizations,
one longitudinal (L, pressure) and one transverse (T, shear)
mode. They are always coupled upon reflections at the boundary. In
this sense, the membrane--like motion is not integrable.

The measured  spectrum in \refFig{fig3} lowest plot  shows evidence of
eigenvalue clusters which are separated by a slowly increasing spacing
in frequency. The first clusters appear only after a gap. At medium to
high frequencies the spacing appears constant and for thin hemispheres
agrees well with the spacing coming from purely T-polarized orbits.
\begin{figure}[htp]
\begin{center}
\includegraphics[width=80mm,height=8cm]{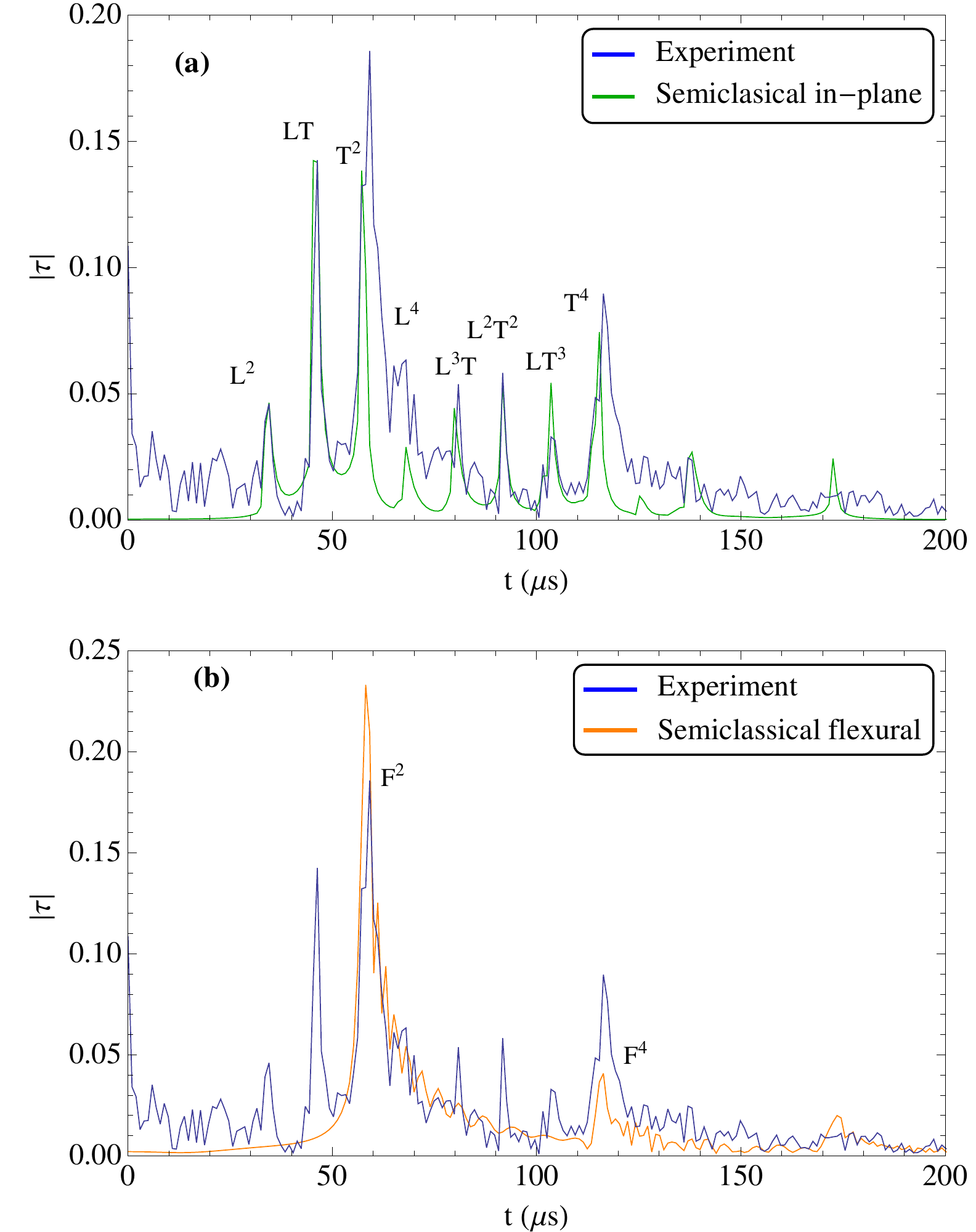}
\end{center}
\caption{(Color online) Time spectrum of spectral fluctuations containing {\bf L}ongitudinal, {\bf T}ransverse  (a) and {\bf F}lexural segments of orbits (b).}
  \label{fig5}
\end{figure}
Furthermore, the extracted discrete spectrum show fluctuations well
described by certain time periods, see Fig.~\ref{fig5}, corresponding
to fixed spacings in frequency which we attribute to periodic orbits
in the following. From shell theory, we therefore focus on the
in-plane deformation field for which dispersion is linear and results
in orbit actions linear in frequency. We shall also touch briefly on
the flexural mode as its dispersion approaches linear at the higher
frequencies probed in our experiment.

The in-plane theory results from ignoring the flexural motion in the
Kirchoff-Love shell theory \cite{kra67,nio88} and leads  
 to 
\beq \frac{E }{2} \left[{ (1+\nu)} \, \Delta
  \mathbf{u} + {(1-\nu)} \, \nabla \nabla \cdot \mathbf{u} \right] +
\rho \omega^2 \mathbf{u} = 0 \eeq 
when dropping curvature terms in the ray limit.  The constants $E$ and
$\nu$ are the elastic modulus of extension and the ratio of Poisson
corresponding to the plane stress approximation used in plates
\cite{kra67}. Up to further curvature terms, this curved version of
Navier-Cauchy's elastic equation is rewritten by re-expressing the
covariant vector laplacean $\nabla_a \nabla^a $ in terms of
two-dimensional curls and a gradient on a divergence \cite{bgv04}.
Consequently, the elastic field can be decomposed in a gradient field
and a curl field corresponding to longitudinal and transverse
polarization each satisfying a curved scalar Helmholtz equation:
\beq
(\Delta  + z_j^2) u^{(j)} = 0 
\eeq
with $j = L,T$ a polarization index in the following and $z_j = k_j R$
dimensionless wave numbers obeying $z_T = \kappa z_L$ with
$\kappa=c_L/c_T$ the ratio of propagation speeds in plates.  On the
sphere, deformations are therefore expanded using Legendre functions
$\{ P,Q \}_n^m(\cos \theta) \exp(i m \phi)$ with $n=l_j$ a
polarization dependent angular momentum obeying $z_j^2 = l_j (l_j+1)$.
We impose free boundary conditions: the integrated
stress tensor across the thickness of the shell vanishes at a
boundary. Thus, the normal components of the stress resultant
vanishes: $N_{\hat{\theta} a}=0$ at $\theta=\pi/2$ with
$a=\hat{\theta},\hat{\phi}$ \cite{kra67}.  The spectrum is then found
by a two-by-two determinantal condition.

As in the scalar problem \cite{son08} we use the method of scattering
quantization \cite{smilanskyCourse} and consider the dynamics of a ray
of polarization $j$ in the angular momentum variable $l_j$ evolving
under the condition that $L_z$ is constant. By elementary calculations
we find this conservation law to be equivalent to the laws of
reflection and refraction for the ray. For the derivation of a trace
formula we therefore sum over $m\equiv L_z=l_L \sin \theta$ with
$\theta$ the incidence angle with respect to the normal.  Except for
the increase in complexity, we find a very similar result as in
\cite{son08}: there is a scattering matrix composed of a propagation
part and a reflection/refraction part.

The propagation part over a single great arc of the hemisphere is
found to be $\exp(-i \pi l_j)$ with $l_j \approx z_j +1/2$ with the
$1/2$ related to the caustic phase shift.  The free propagation over
the sphere is then independent of $m$ and no saddle point integration
is needed as in the case of an opening angle different from $\pi/2$.

The total $m$-dependence resides only in the reflection coefficients
$\alpha_q$ used for constructing an orbit $q$.  For large $l$ each
reflection coefficient is asymptotic to the classical reflection
coefficient and a slowly varying function of $m$, so the sum over $m$
is well approximated by an integral and yields
\bea
\tr \alpha_q &\approx& \int_{-l_L}^{l_L} dm \,
\alpha_q\left(\frac{m}{l_L}\right) \\ \nonumber &=& l_L \,
\int_{-\frac{\pi}{2}}^{\frac{\pi}{2}} d\theta \, \cos \theta \,
\mathcal{\alpha}_q(\theta)\equiv l_L \, \overline{\mathcal{A}_q} \, . 
\eea
The final result of our calculations is that the oscillating number of
states $d\tilde{N}$ in a frequency interval $df$ coming from orbits is
given to leading order as
\beq
\label{eq:dosFamily}
d\tilde{N}=\tilde{\rho}_{IP}(f) df \approx l_L \, \sum_{p^r}  {\overline{\mathcal{A}^r_p}}\, \cos(\pi r l_p) dl_p \, ,
\eeq
where $p^r$ is a closed orbit containing an $r'$th repeat of a prime
ray sequence $p$ and $l_p = n_L l_L + n_T l_T$ with $n_L,n_T$
integers.  We find it useful to group orbits with identical actions
such as $(LT)^2$ and $LLTT$ and denote the family $L^2T^2$.  In
practice ${\overline{\mathcal{A}}}$ for each family is calculated
numerically.  Notably, fluctuations grow linearly with frequency to
leading order.  Rayleigh edge orbits are omitted as their contribution
is of order $f^0$ only.

For completeness we include transverse rays with incidence angles
beyond the critical angle for conversion for which the reflection
coefficient is of unit modulus. Consequently, this part of phase space
for T-polarized orbits leads to clusters of closely grouped states
\cite{SAFAROV}.  Ref.~\cite{SAFAROV} states that \refeq{eq:dosFamily}
should hold in general; here for the first time we present an explicit
study in a ray splitting case. We give a parallel treatment to the
flexural orbits using their corresponding reflection coefficients
\cite{HB} and dispersion relation \cite{gap}.  Here we settle for one
extracted from the numerical solution of a whole spherical shell due
to its larger spectral range.

The simulated spectral fluctuations of the hemisphere used in our
experiment have a period spectrum depicted in \refFig{fig5}: the (a)
spectrum shows clear evidence of in-plane orbits. The inclusion of the
flexural modes in (b) gives trains of peaks with the main peaks close
to those of the transverse. This is because at higher frequencies the
flexural dispersion approaches the linear with the flexural speed
approaching the Rayleigh speed which is just below the transverse
speed. Furthermore, the clusters in the transducer signal we checked
to agree well with the peaks in the flexural fluctuations.

%\begin{figure}[htp]
%\begin{center}
%\includegraphics[width=80mm]{fig6}
%\end{center}
%\caption{(Color online) Time spectrum of spectral fluctuations containing {\bf F}lexural segments of orbits.}
 % \label{fig6}
%\end{figure}

Why is the spectrum for the hemisphere so different? This is
illustrated in \refFig{fig4}. For all opening angles, all periodic
orbits are geodesic polygons, \refFig{fig4} (left) shows the pentagram
as an example. As the opening angle increases, the surface enclosed by
the polygons becomes larger. At $\theta_0=90^\circ$, this surface is
the entire hemisphere. Hence all polygons degenerate to the orbit on
the base of the hemisphere. The diameter orbit shown in \refFig{fig4}
(right) is the exception: there is only one such orbit for
$\theta_0<90^\circ$, because it must go through the north pole. At
$\theta_0=90^\circ$, however, all geodesic lines connecting opposite
points on the base are diameter orbits.  These drastic changes in the
periodic orbit structure reflect the equally drastic  changes
in the spectrum when the opening angle reaches $90^\circ$.  Formula
\refeq{eq:dosFamily} gives the precise mathematical connection.

In conclusion, we measured high--resolution spectra of a family of
open spherical shells and saw clear evidence of in-plane behavior. For
this we developed a trace formula for the density of states which
agree well with experiment.  The amplitudes of the fluctuations were
not entirely as in experiment, still the location of the periods of
the experimental fluctuations agreed well with theory.  Hence, we have
presented a first identification of periodic orbit structures on
curved geometries.  From the simple form of the trace formula we
expect it to be easy to generalize and apply to other wave equations
on the hemisphere.  Thus, when applied to the flexural mode, also the
clusters of the transducer signal are described.

\begin{acknowledgments}
  We thank Sir M.~Berry and E.~Bogomolny who already a long time ago
  suggested shells to us as an important object of study.  TG
  acknowledges support from Deutsche Forschungsgemeinschaft (SFB/TR12,
  ``Symmetries and Universality in Mesoscopic Systems''), NS is
  grateful for support from Det Svenska Vetenskapsr\aa det.
\end{acknowledgments}


\begin{thebibliography}{99}

\bibitem{gut90}       M.C. Gutzwiller,
                      {\it Chaos in Classical and Quantum Mechanics},
                      Springer, New York (1990)
\bibitem{bra97}       M. Brack and R.K. Bhaduri,
                      {\it Semiclassical Physics},
                      Frontiers in Physics, Vol. {\bf 96},
                      Addison-Wesley, Reading (1997)
\bibitem{sto99}       H.J. St\"ockmann,
                      {\it Quantum Chaos --- An Introduction},
                      Cambridge University Press, Cambridge (1999)
\bibitem{haa01}       F. Haake, 
                      {\it Quantum Signatures of Chaos}, 2nd ed., 
                      Springer, Heidelberg (2001)





\bibitem{heu07}       S. Heusler, S. M\"uller, A. Altland, P. Braun, and F. Haake,
                      Phys. Rev. Lett. {\bf 98}, 044103 (2007)
\bibitem{fri89}       H. Friedrich and D. Wintgen,
                      Phys. Rep. {\bf 183}, 37 (1989)
\bibitem{win92}       D. Wintgen, K. Richter, and G. Tanner,
                      Chaos {\bf 2}, 19 (1992)
\bibitem{mc2b}        H.J. St\"ockmann and J. Stein,
                      Phys. Rev. Lett. {\bf 64}, 2215 (1990);
                      H.D. Gr\"af, H.L. Harney, H. Lengeler, C.H. Lewenkopf, 
                      C. Rangacharyulu, A. Richter, P. Schardt,
                      and H.A. Weidenm\"uller,
                      Phys. Rev. Lett. {\bf 69}, 1296 (1992)
\bibitem{ac3b}        R.L. Weaver,
                      J. Acoust. Soc. Am. {\bf 85}, 1005 (1989);
                      C. Ellegaard, T. Guhr, K. Lindemann, H.Q. Lorensen,
                      J. Nyg\aa rd, and M. Oxborrow,
                      Phys. Rev. Lett. {\bf 75}, 1546 (1995)
\bibitem{ber99}       P. Bertelsen, C. Ellegaard, T. Guhr, M. Oxborrow,
                      and  K. Schaadt,
                      Phys. Rev. Lett. {\bf 83}, 2171 (1999)
\bibitem{wea00}       R.L. Weaver and O. Lobkis,
                      J. Sound Vib. {\bf 231}, 1111 (2000)
\bibitem{sch03}       K. Schaadt, T. Guhr, C. Ellegaard, and M. Oxborrow,
                      Phys. Rev. {\bf E68}, 036205 (2003)
\bibitem{gui06}       L. Guti\'errez, A. Diaz--de--Anda, J. Flores, 
                      R.A. M\'endez--S\'anchez, G. Monsivais, 
                      and A. Morales, 
                      Phys. Rev. Lett. {\bf 97}, 114301 (2006)


\bibitem{kra67}       H. Kraus,
                      {\it Thin Elastic Shells},
                      Wiley, New York (1967)

\bibitem{nio88}       F.I. Niordson,
                      Int. J. Solids Structures {\bf 24}, 947 (1988)




\bibitem{dah98}       F.A. Dahlen and J. Tromp,
                      {\it Theoretical Global Seismology},
                      Princeton University Press, Princeton (1998)


\bibitem{del94}       D. Delande, D. Sornette, and R.L. Weaver,
                      J. Acoust. Soc. Am. {\bf 96}, 1873 (1994)                 


\bibitem{son08}       N. S\o ndergaard and T. Guhr,
                      J. Phys. {\bf A41}, 075309 (2008)


\bibitem{bgv04} N. Berline, E. Getzler, and M. Vergne {\it Heat Kernels and Dirac Operators}, Springer, New York (2004)




\bibitem{smilanskyCourse} U. Smilansky  in {\em
                        Les-Houches Summer School on mesoscopic
                        quantum physics}, edited by E. Akkermans et
                      al., North-Holland (1994)


\bibitem{SAFAROV}     Y. Safarov  and D. Vassiliev    
                      in {\em Spectral Theory of Operators}, 
                      edited by S.Gindikin, 
                      AMS Translations,  ser.2, {\bf 150} (1992)


\bibitem{HB}          E. Bogomolny  and E. Hugues  
                      { Phys. Rev.} {\bf E57} 5404 (1998)



\bibitem{gap}         A.D. Pierce,
                      Structural Acoustics, ASME, NCA Vol.~12,
                      AMD Vol. 128, 195 (1991);
                      A.D. Norris and D.A. Rebinsky,  
                      J. Vib. Acoust.                     
                      {\bf 116}, 457 (1994) 


\end{thebibliography}
\end{document}